\documentclass{elsarticle}
\usepackage{hyperref}
\usepackage{graphicx}  
\usepackage[utf8]{inputenc}
\usepackage[T1]{fontenc}
\usepackage{siunitx}[=v2]
\let\qty\SI

\journal{Nucl. Instrum. Methods Phys. Res. A}
\begin{document}%
\begin{frontmatter}
\title{The MAPS foil}

\author[1]{S.~Beolé}
\author[2]{F.~Carnesecchi}
\author[3]{G.~Contin}
\author[2]{R.~de Oliveira}
\author[2]{A.~di Mauro}
\author[2]{S.~Ferry}
\author[2]{H.~Hillemanns}
\author[2]{A.~Junique}
\author[2]{A.~Kluge}
\author[2,4]{L.~Lautner}
\author[2]{M.~Mager\corref{cor}}
\cortext[cor]{Corresponding author: Email: \url{Magnus.Mager@cern.ch}}
\author[2]{B.~Mehl}
\author[2,5]{K.~Rebane}
\author[2]{F.~Reidt}
\author[2,4]{I.~Sanna}
\author[2]{M.~Šuljić}
\author[6]{A.~Yüncü}
\address[1]{University and INFN Torino, Torino, Italy}
\address[2]{CERN, Geneva, Switzerland}
\address[3]{University and INFN Trieste, Trieste, Italy}
\address[4]{Technical University of Munich, Munich, Germany}
\address[5]{Tallinn University of Technology, Tallinn, Estonia}
\address[6]{University of Heidelberg, Heidelberg, Germany}

\begin{abstract}
We present a method of embedding a Monolithic Active Pixel Sensor~(MAPS) into a flexible printed circuit board~(FPC) and its interconnection by means of through-hole copper plating. The resulting assembly, baptised ``MAPS foil'', is a flexible, light, protected, and fully integrated detector module. By using widely available printed circuit board manufacturing techniques, the production of these devices can be scaled easily in size and volume, making it a compelling candidate for future large-scale applications.

A first series of prototypes that embed the ALPIDE chip has been produced, functionally tested, and shown to be working.
\end{abstract}


\end{frontmatter}

\section{Introduction}
Monolithic Active Pixel Sensors (MAPS) have a number of distinctive properties, which make them the technology of choice for numerous tracking and vertex detectors, where low material budgets and high intrinsic resolutions are key (e.g.~\cite{STAR:PIXEL,ITS2:TDR,ITS3:LoI,ALICE3:LoI,sPHENIX,CBM,NICA:MAPS,LUXE,mu3e:TDR,Belle2}). The ALPIDE chip~\cite{ALPIDE}, which was developed for the new ALICE Inner Tracking System (``ITS2''), for instance, achieves an intrinsic spatial resolution of~\qty{5}{\um} and has a nominal thickness of only~\qty{50}{\um}, corresponding to~\qty{0.05}{\percent} of radiation length ($X_0$). The ITS2 covers an area of~\qty{10}{\metre\squared}; it has recently been installed and is taking data at LHC, marking the first large-scale application of MAPS in high energy physics. A~\qty{60}{\metre\squared}~tracker is envisaged for the ALICE3 detector, targeting installation during the LHC Long Shutdown~4~(2033--34), and asking for new module concepts that can be industrialised, while at the same time meeting stringent material budget constraints~\cite{ALICE3:LoI}.

The R\&D for replacing the inner-most layers of ITS2 by new MAPS layers (ITS3 project~\cite{ITS3:LoI}) based on bent, wafer-scale sensors is currently ongoing. Through a reduction of  the material budget by a factor of seven with respect to ITS2 and moving closer to the interaction point, ITS3 aims to achieve unprecedented position resolution figures. First results demonstrate the feasibility of bent MAPS~\cite{ITS3:bent}, paving the way to new, optimised detector geometries. Within this R\&D for bending thin silicon, the here-presented technique to both protect and interconnect sensors was conceived, making the technology potentially even more versatile while adding only a small amount of material.

Current lowest mass detector assemblies are typically based on thin chips that are glued onto thin printed circuit boards and interconnected e.g.~with wire bonds or SpTAB bonding (see e.g.~\cite{SpTAB}). These assemblies are then held in flat position by external support mechanics or stiffeners.

In this paper, we propose, and demonstrate, a method to fully embed \qty{50}{\um}-thin MAPS into polyimide flexible printed circuit boards. The resulting assembly does not only protect the silicon chip, but also provides the electrical interconnection. Furthermore, the assembly remains flexible and protects the chips from breakage when bending them. Excluding the copper layer, our prototype has a material budget of around \qty{0.1}{\percent}~$X_0$, only twice that of the chip alone. Future optimisations of the process can potentially lower this further if required.  The fill-factor of the copper layer will be subject to optimisation and hence its relative contribution to the total material budget will be highly application-dependent.

The process is inspired by previous works~\cite{SERVIETTE, DulinskiFEE2014}, but differs in the choice of interconnection metallurgy and does not require a chip-level redistribution layer, simplifying the procedure and, importantly, making it compatible with industry-standard manufacturing techniques.

Our process can easily be extended to larger areas, integrating and interconnecting several chips. It can also be considered as a means of adding a coarse routing layer to large, stitched chips, for instance for improved power distribution.

The resulting assembly is a flexible, highly integrated, thin, and yet robust vertexing layer, that can be scaled to large surfaces.

\section{The MAPS foil concept}
The basic idea of the MAPS foil is to embed MAPS sensors, either single or multiple ones, into flexible printed circuit boards. A sandwich of polyimide films, glue and sensor is made, and the copper traces are connected to the sensor by through-hole metallisation.

The assembly provides three main features at once:
\begin{itemize}
    \item\textbf{Mechanical support:} the polyimide serves as a support structure to which precise mechanical alignment holes can be added; in addition, a number of chips can be placed next to each other to form a multi-chip module
    \item\textbf{Mechanical protection:} the polyimide protects the silicon; it can be manipulated by bare hands easily
    \item\textbf{Electrical interconnection:} the top polyimide film can be copper-clad and structured to form electrical traces, like in ordinary printed circuit boards; they can be interconnected to the chip pads, by opening holes and metallising them
\end{itemize}

The foil moreover remains flexible (as long as the sensor is\footnote{Despite the different layers of material, preliminary tests show that sensors can easily be bent to $R=\qty{30}{\mm}$ (and very likely much further). More details will be subject of a future publication.}) and can be shaped into e.g.~cylindrical shapes. It can indeed prevent early breakage by smoothing out the stress distribution over the silicon.

\section{Process}
\begin{figure}
    \centering
    \includegraphics[width=\textwidth]{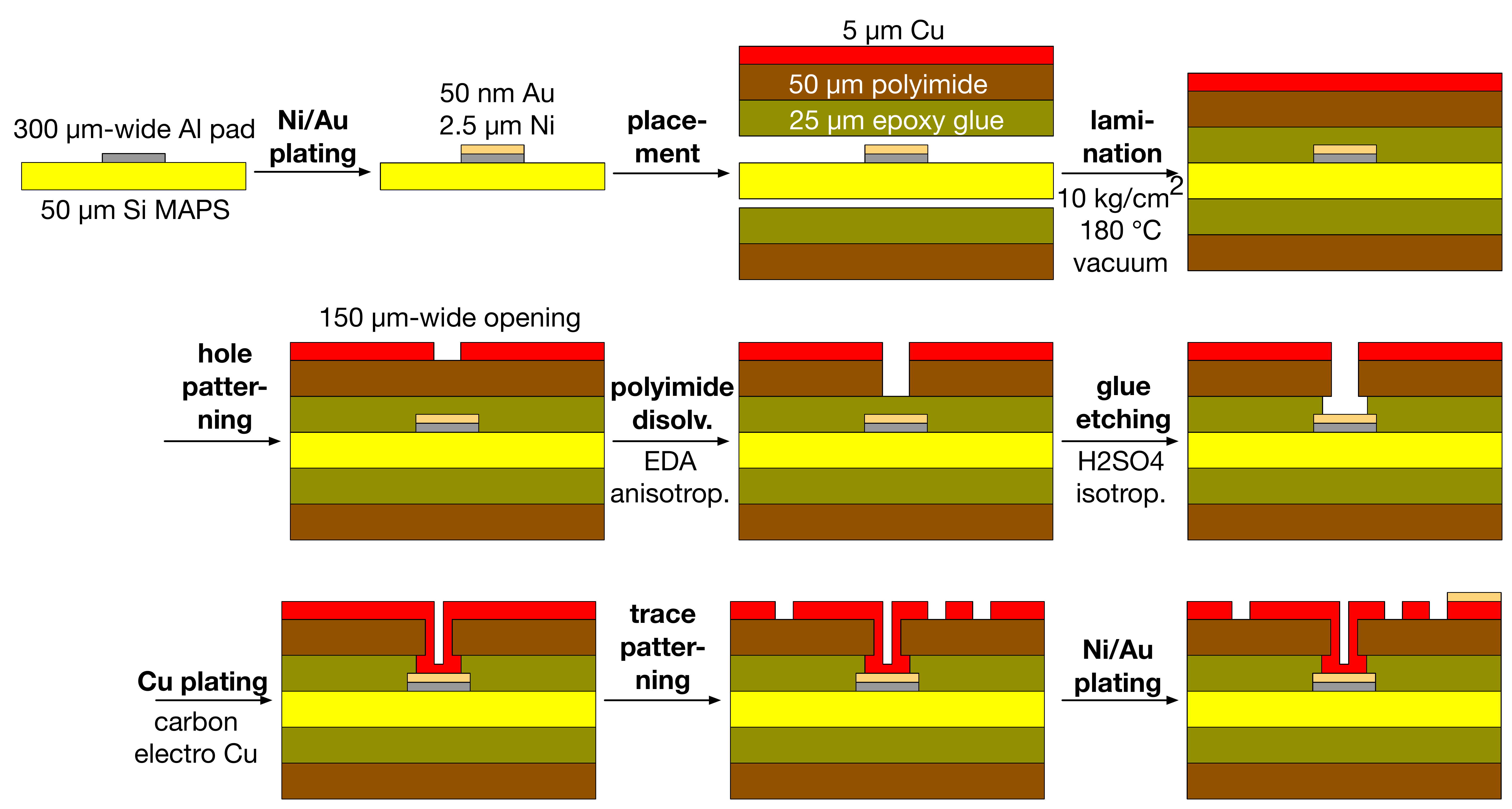}
    \caption{The MAPS~foil production process steps (from top left to bottom right).}
    \label{fig:process}
\end{figure}
The creation of the MAPS foil consists of two main stages: the embedding of the chip(s) between two polyimide films, and the electrical interconnection. The process, which is similar to the production of multi-layer printed circuit boards, is depicted in Fig.~\ref{fig:process} and discussed hereafter. The mentioned material thicknesses are those used for the production of the first prototype and can be further optimised.

\subsection{Chip preparation}
The interconnection to the chip is based on metallisation, in a fashion similar to via plating in ordinary PCB manufacturing. For the metallurgy to work, the aluminium pads of the chip are first nickel-gold plated. A~\qty{2.5}{\um}-thick nickel layer is deposited with a~\qty{50}{\nm}-thick gold finish.

\subsection{Embedding}
To tie the chip(s) to the polyimide, a thin glue layer is applied to the latter. The glue stems from an FR4 pre-preg\footnote{Ditron PRG-EP-84} that is first applied to the polyimide film, and from which the fibres are removed afterwards. The polyimide-glue-chip-glue-polyimide sandwich is then compressed and heated to cure the glue.

\subsection{Interconnection}
For the interconnection, a copper-clad polyimide is used when embedding the chip. In the next steps a hole is opened over the pad positions, subsequently in the copper layer, the polyimide and the glue. This is done in a wet-chemical fashion, with different chemicals for the different layers\footnote{The process builds largely on the experience that was gained with GEM foils~\cite{GEM}.}. The copper etching is done by applying a UV mask, and the subsequent steps are self-aligned, i.e.~they use the layers above as masks. Since the chip can be seen from below (there is no copper cladding underneath the chip), the alignment of the first mask can be done quite accurately, certainly to the required level of the order of~\qty{50}{\um}, which is given by the attainable structure size and the dimensions of the chip pads. Once the hole is opened, copper electro-plating is performed, creating the metal interconnection between the pad and the copper layer.

\subsection{Trace patterning}
The final steps consist in etching the traces on the copper (similar fashion as the hole opening in the copper), and the nickel-gold plating of contacts for interconnection of the foil to other devices.

\section{Prototype}
\begin{figure}
    \centering
    \includegraphics[width=0.75\textwidth]{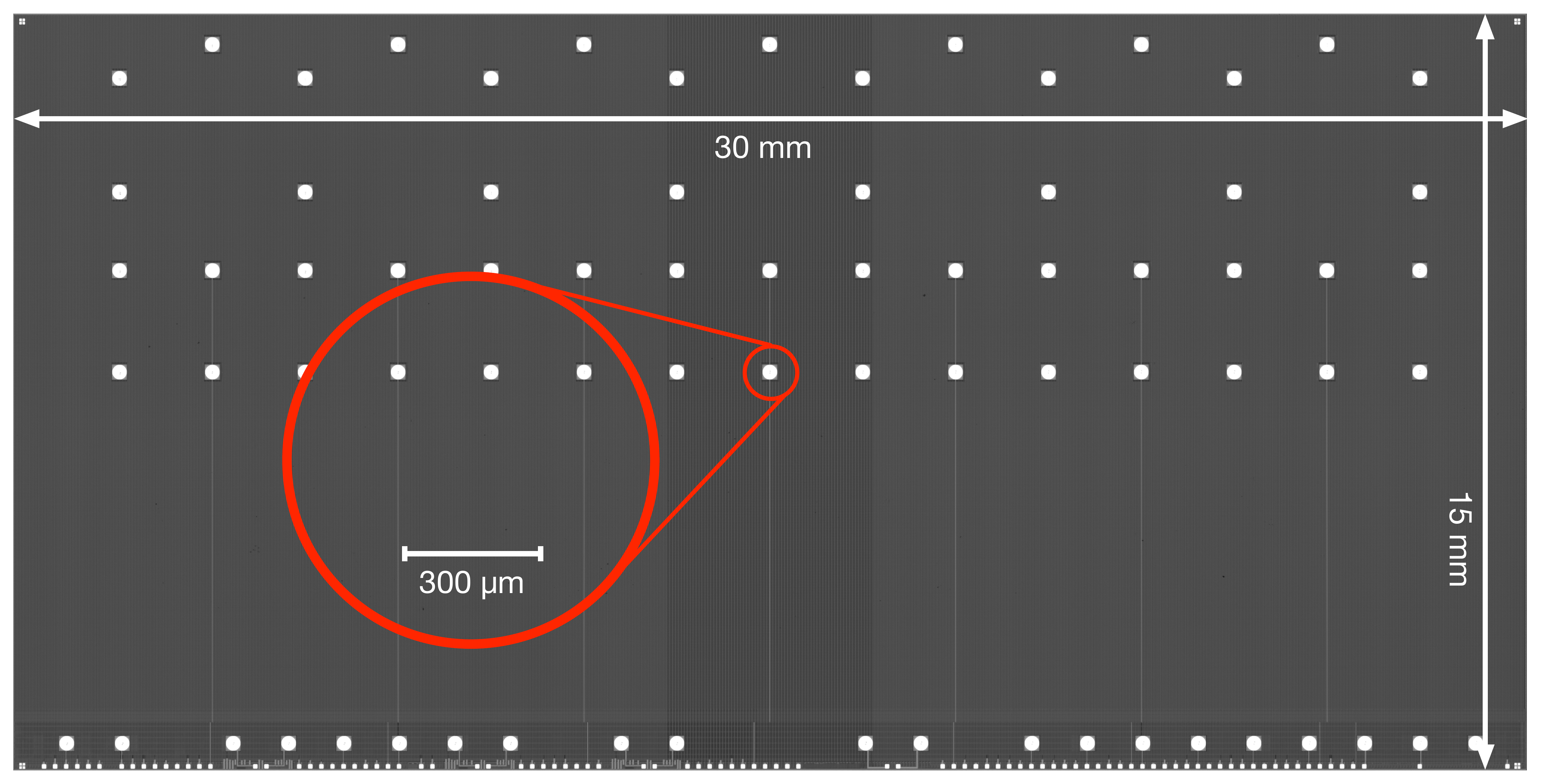}
    \caption{Photograph of the ALPIDE chip, displaying the pattern of the large pads used for interconnection, zooming on one of them.}
    \label{fig:ALPIDE}
\end{figure}
A first prototype was realised using ALPIDE sensors developed for the ITS2. ALPIDEs are \mbox{\SI{30 x 15}{\mm}}-large MAPS fabricated in the TowerJazz \qty{180}{\nm} CMOS Imaging Technology~\cite{TJ180}, that are readily thinned down to approximately~\qty{50}{\um}. They feature about~\num{60} large, \qty{300}{\um}-diameter, pads that are distributed over the chip surface for interconnection to detector modules\footnote{Within the ITS2, ALPIDEs are interconnected using wire-bonds through openings in a standard flexible printed circuit board that is placed over the chips, though different schemes were tried during the R\&D phase, notably including laser soldering~\cite{ITS2:TDR}.}, making them suited for the process explained here~(cf.~Fig.~\ref{fig:ALPIDE}).

\begin{figure}
    \centering
    \includegraphics[width=0.75\textwidth]{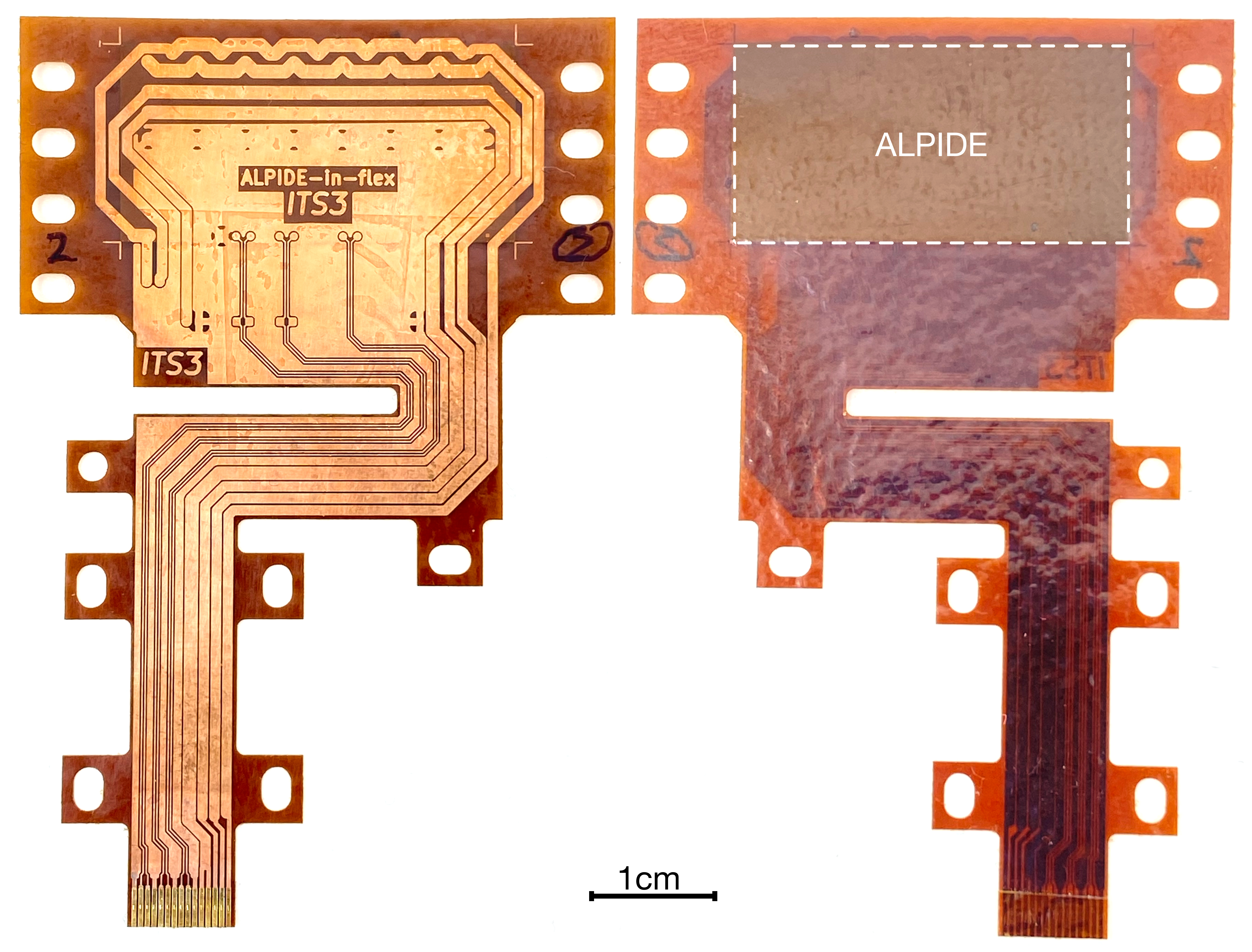}
    \caption{Photographs (front and back) of the assembly. The ALPIDE is embedded at the top part (annotated on the right).}
    \label{fig:assembly}
\end{figure}
Figure~\ref{fig:assembly} shows what the prototype assembly looks like. It is a single-sided circuit board, that can be connected to existing readout hardware developed within the ALICE ITS MAPS R\&D~\cite{DAQ}.

Five of these assemblies were fabricated and electrically tested for possible malfunctions of the sensors\footnote{The sensors were tested systematically beforehand by means of probe testing.}. None was showing evidence for breaking the sensor, while two had a slight misalignment of the pads and holes, leading to non-working interconnections. This was spotted already during manufacturing, and we do not consider this a principle problem.

\begin{figure}
    \centering
    \includegraphics[width=\textwidth]{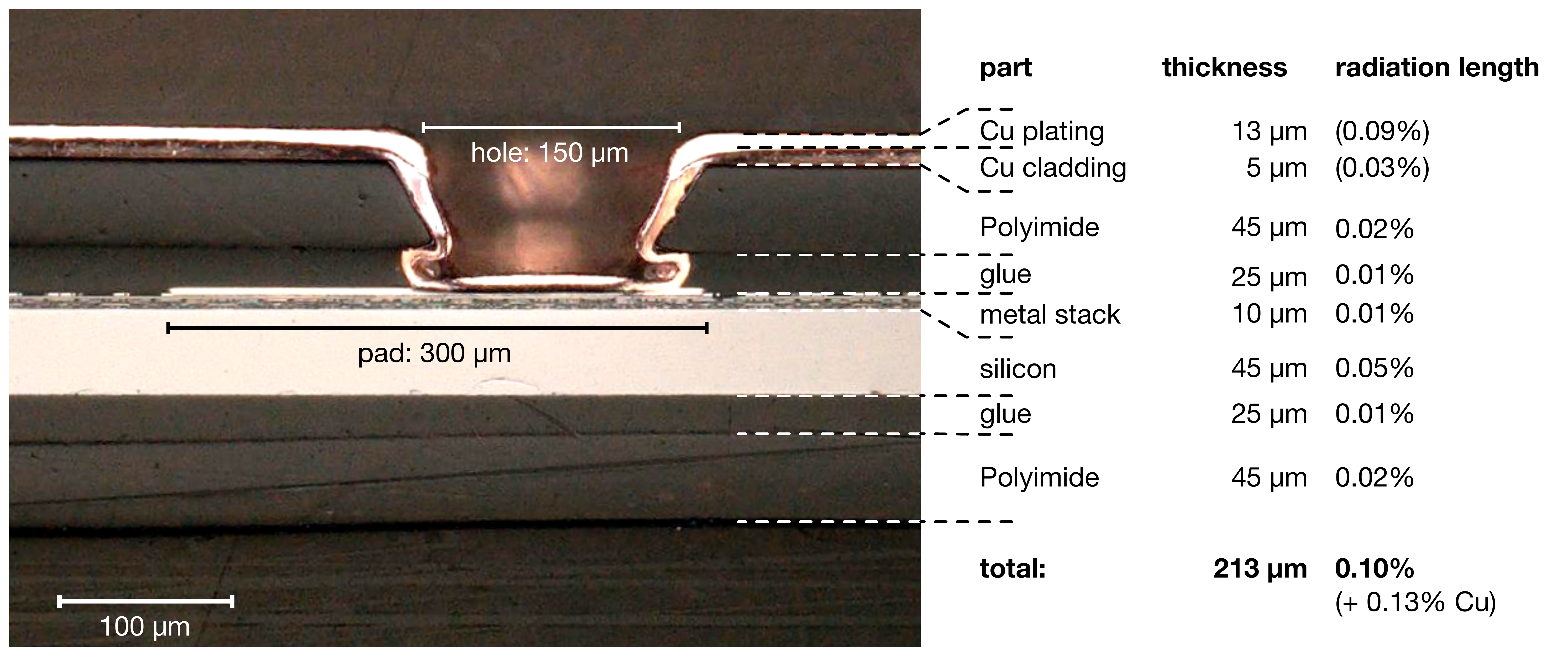}
    \caption{Cross-section of one interconnection point and associated material budget breakdown. Radiation lengths from~\cite{PDG}, and assuming a generic value of \qty{35.5}{\cm} for the epoxy glue.}
    \label{fig:xsection}
\end{figure}
Figure~\ref{fig:xsection} shows the cross-section through one of the interconnections. The composition of the foil, the different characteristic geometries obtained from polyimide and glue openings, and the deposited copper plating that makes a solid contact are clearly visible. A material budget breakdown is given alongside, showing that the ``extra'' material accounts for roughly as much as the sensor itself. The copper layer is not included here, as its contribution will scale with the fill-factor of the traces as well as with the total thickness allowed for each specific application.

\begin{figure}
    \centering
    \includegraphics[width=\textwidth]{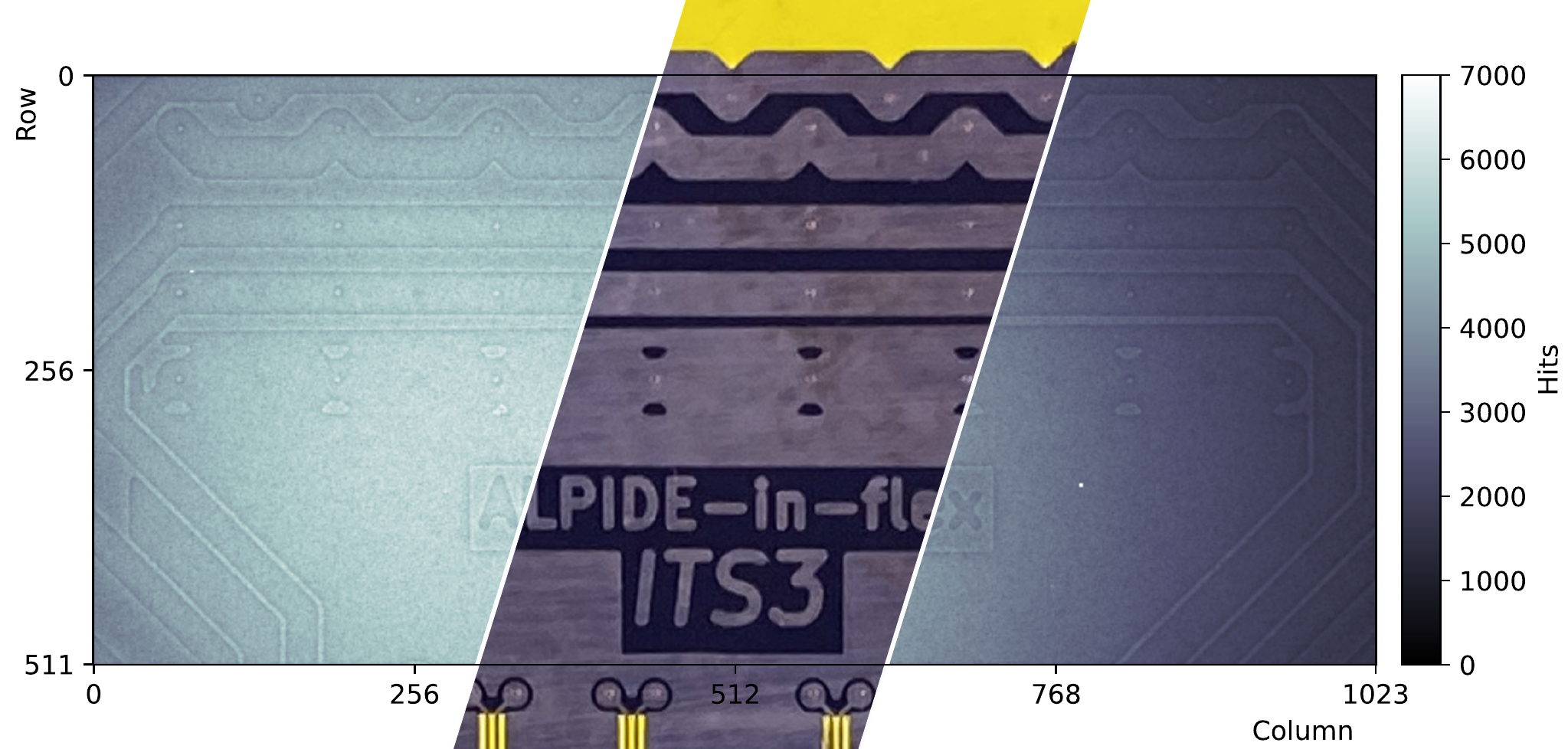}
    \caption{Beta scattering image, self-portrait, of the assembly as acquired by the embedded chip, while being exposed to electrons from a Sr-90 source. The central part is overlaid with a photograph of the assembly to guide the eye. The drop of hit frequency towards the right and edges of the pixel array is due to the positioning of the source.}
    \label{fig:beta}
\end{figure}

Three devices were successfully tested for functionality, including their response to an Sr-90~beta~source, which created beta scattering images of the copper layers as shown in Fig.~\ref{fig:beta}, nicely demonstrating the functioning of the assembly.

\section{Discussion and relation to other methods}
We would like to point out a few aspects that are key to our method, and make it distinct from other approaches:
\begin{itemize}
    \item\textbf{Subtractive manufacturing:}
    The application of a solid glue layer and creation of the holes afterwards in a subtractive manner has an advantage over segmented gluing (e.g.~by glue dispensing), since its initial application can be controlled more easily.
    \item\textbf{Symmetry:}
    The polyimide-chip-polyimide symmetry of the assembly reduces deformation after the gluing due to a mismatch of thermal expansion coefficients of chip, glue and polyimide.
    \item\textbf{Copper:}
    Using copper as metallisation material makes the method compatible with standard PCB/FPC manufacturing. An alternative approach using aluminium is however being investigated, since it allows to reduce the material budget. It would have the advantage of connecting aluminium pads to aluminium traces, but this requires custom equipment. There are, however, important difficulties when working with aluminium, which may motivate the use of copper. Aluminium is more fragile, tending to break more easily, effectively reducing the integration density. It also needs to be sputtered (not electroplated), which makes the creation of metallised holes much more difficult.
    \item\textbf{Interconnection method:}
    The interconnection by metallisation has a coarser feature size than, for instance, wire bonding. It has the advantage of a solid and robust interconnection, without the need of extra protection. In principle, wire bonds through larger openings could coexist if a higher connection density is needed.
\end{itemize}

\section{Summary \& outlook}
We demonstrated a method to construct detector modules by embedding sensors in polyimide films. The resulting first ``MAPS~foils'' have outstanding properties, potentially allowing the production of large detector systems in an industrial fashion.

This first successful trial also sparked a number of activities aiming at the construction of multi-chip modules, embedding of wafer-scale chips, as well as using chips produced in a  \qty{65}{\nm} process, which is the target technology node for the ITS3. In addition, investigations of using a laser assisted process for the hole formation and using aluminium instead of copper are ongoing. Finally, the robustness, including the ageing of the assemblies, is being studied.

MAPS foils are a novel way to solve the module integration problem, bridging semiconductor sensors to macroscopic detector elements in an integrated fashion. They are protected, versatile, and flexible, yet only adding a minimal amount of material to the bare sensors. This is an appealing combination for many applications that require minimal material budgets.

\section*{Acknowledgements}
M.~Mager feels indebted to the late W.~Dulinski for his inspiring demonstration of a very similar prototype at the Front End Electronics Workshop at the Argonne National Laboratory in May~2014~\cite{DulinskiFEE2014}.

\bibliography{8references}

\end{document}